\documentclass[%
 reprint,
 superscriptaddress,
 preprintnumbers,
 nofootinbib,
 amsmath,amssymb,
 aps,
 prl,
]{revtex4-1}
\usepackage{psfrag,slashed,cancel,array,graphicx,todonotes,hyperref}
\usepackage[justification=justified]{subcaption}
\usepackage[labelfont=bf,labelsep=quad,justification=justified]{caption}
\usepackage[utf8]{inputenc}
\usepackage{mathtools}
\usepackage{mathrsfs}
\usepackage{multirow}
\usepackage{booktabs} 
\usepackage[normalem]{ulem}
\hypersetup{pdftitle={},pdfcreator={},linkcolor=[rgb]{0.15,0.35,0.75},colorlinks=true,citecolor=[rgb]{0.675,0,0.2},urlcolor=[rgb]{0.15,0.35,0.65}}

\def\beq{\begin{equation}}
\def\eeq{\end{equation}}
\def\bsp#1\esp{\begin{split}#1\end{split}}

\AtBeginDocument{
\heavyrulewidth=.08em
\lightrulewidth=.05em
\cmidrulewidth=.03em
\belowrulesep=.65ex
\belowbottomsep=0pt
\aboverulesep=.4ex
\abovetopsep=0pt
\cmidrulesep=\doublerulesep
\cmidrulekern=.5em
\defaultaddspace=.5em
}

\def\eps{\epsilon}

\begin{document}

\preprint{CERN-TH-2019-052}
\preprint{CP3-19-19}
\preprint{MIT-CTP/5115}
\preprint{SLAC-PUB-17425}

\title{Higgs production in bottom-quark fusion to third order in the strong coupling}

\author{Claude Duhr}
\affiliation{Theoretical Physics Department, CERN,
CH-1211 Geneva 23, Switzerland.}
\affiliation{Centre for Cosmology, Particle Physics and Phenomenology (CP3),
UCLouvain, Chemin du Cyclotron 2, 1348 Louvain-La-Neuve, Belgium.}
\author{Falko Dulat}
\affiliation{SLAC National Accelerator Laboratory, Stanford University, Stanford, CA 94039, USA.}
\author{Bernhard Mistlberger}
\affiliation{Center for Theoretical Physics, Massachusetts Institute of Technology, Cambridge, MA 02139, USA.}

\begin{abstract}
We present the inclusive cross section at next-to-next-to-next-to-leading order (N$^3$LO) in perturbative QCD for the production of a Higgs boson via bottom-quark fusion. 
We employ the five-flavour scheme, treating the bottom quark as a massless parton while retaining a non-vanishing Yukawa coupling to the Higgs boson. 
We find that the dependence of the hadronic cross section on the renormalisation and factorisation scales is substantially reduced. 
For judicious choices of the scales the perturbative expansion of the cross section shows a convergent behaviour. 
We present results for the N$^3$LO cross section at various collider energies.
In comparison to the cross section obtained from the Santander-matching of the four and five-flavour schemes we predict a slightly higher cross section, though the two predictions are consistent within theoretical uncertainties.
\end{abstract}

\maketitle

With the discovery of the Higgs boson by the experiments at the Large Hadron Collider (LHC) at CERN~\cite{Aad:2012tfa,Chatrchyan:2012xdj} particle physics has entered a new era. 
The last missing degree of freedom of the Standard Model (SM) of particle physics has  been established, making the SM a fully predictive theory without any free parameters. 
For the first time we can now directly measure the properties of a fundamental scalar boson. 
The remarkable performance of the LHC during Run II brings the Yukawa
interactions
- the interaction among fundamental fermions and the Higgs boson - 
within reach of being probed by the experiments.
Clearly, exploring the Higgs sector and testing our understanding of the fundamental interactions of nature is one of the main tasks for the LHC.

Since in the SM the coupling strength of the Higgs boson to fermions is proportional to the fermion mass, prospects for coupling measurements are most promising for third-generation matter, i.e., bottom and top quarks and $\tau$ leptons. 
Several models of new physics, including the Minimal Supersymmetric Standard Model, predict enhanced couplings of the Higgs boson to bottom quarks. 
Consequently, studying the Yukawa coupling to bottom quarks is particularly interesting.
The decay of a Higgs boson to pairs of bottom quarks is known to high orders in
perturbation theory~\cite{Braaten:1980yq, Surguladze:1994gc, Kataev:1993be, Larin:1995sq, Chetyrkin:1995pd, Chetyrkin:1996sr, Baikov:2005rw, Anastasiou:2011qx, DelDuca:2015zqa, Caola:2017xuq, Mondini:2019gid, Primo:2018zby}.
While this decay channel benefits from a large
branching fraction, it is overwhelmed by QCD background and thus challenging to observe~\cite{Aaboud:2018zhk,Sirunyan:2018kst}.
A viable alternative consists in probing the couplings of the Higgs boson to bottom quarks via the production of a Higgs through fusion of a $b\bar{b}$ pair.

There exist two formally equivalent descriptions of the cross section for the production of a Higgs boson from the fusion of bottom quarks. 
In the four-flavour scheme (4FS), the bottom quark is treated as a massive quark, which decouples from the evolution of the strong coupling constant and the parton density functions (PDFs) described by the Dokshitzer-Gribov-Lipatov-Altarelli-Parisi (DGLAP) equation. 
At leading order (LO) in perturbative QCD in the 4FS  bottom quarks are produced from gluon splitting. 
Since the bottom quarks are treated as massive, the gluon splitting is free of collinear divergences, but produces logarithms $\log m_b^2/Q^2$ at every order in perturbation theory, where $Q$ is a characteristic scale of the hard process. 
These logarithms may become large, thereby spoiling the convergence of the perturbative series and requiring resummation. 
In the five-flavour-scheme (5FS), instead, the bottom quark is treated as a massless parton and is included in the evolution of the strong coupling and the PDFs. 
The collinear logarithms are then de facto resummed into the PDF evolution, thereby avoiding the appearance of large logarithms order-by-order in perturbation theory. 
On the other hand, the inclusive cross section in the 5FS neglects power-suppressed terms of order $m_b^2/Q^2$, which are automatically included in the 4FS where all mass effects are taken into account. 
While both schemes are formally equivalent non-perturbatively, the truncation of the perturbative series is necessary for practical calculations and introduces a scheme dependence of the cross section.
\begin{center}
\begin{table*}[!t]
\begin{center}
\begin{tabular}{c|c|c|c|c}
\hline\hline
 & \includegraphics{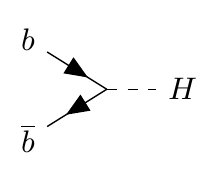} & \includegraphics{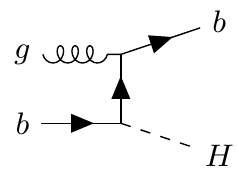} & \includegraphics{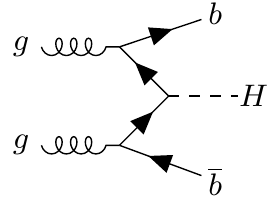} & \includegraphics{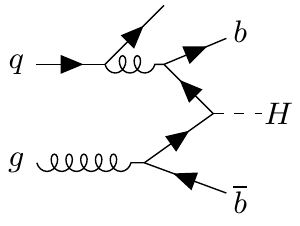}\\
 \hline &&&&\\
  $\quad$4FS$\quad$ & -- & -- & LO & NLO \\
 $\quad$5FS$\quad$ & LO & NLO & NNLO & N$^3$LO \\
 Partonic channels (5FS) & $b\bar{b}$ & $b\bar{b},bg$ & $b\bar{b},bg,bb,bq,b\bar{q},gg,q\bar{q}$ & $b\bar{b},bg,bb,bq,b\bar{q},gg,q\bar{q},qg$\\
 &&&&\\
  \hline\hline
\end{tabular}
\end{center}
\caption{\label{fig:diagrams} \centerlast Representative diagrams contributing at different orders in perturbation theory in the 4FS and 5FS. The last line summarises the partonic channels in the 5FS. Channels related by charge conjugation are not shown explicitly and $q$ denotes a light quark that does not couple directly to the Higgs boson. The partonic channels in the 4FS are obtained by ignoring initial states involving a bottom quark.}
 \end{table*}
 \end{center}
 
Tab.~\ref{fig:diagrams} displays representative Feynman diagrams contributing in the 4FS and 5FS respectively. 
At LO, the 5FS is described by a two-to-one process. 
The simple structure of the LO process makes it possible to compute the inclusive cross section to high orders in perturbation theory, and both next-to-leading order (NLO)~\cite{Dicus:1998hs,Balazs:1998sb} and next-to-next-to-leading order (NNLO)~\cite{Harlander:2003ai} results are available. 
Computations in the 4FS are technically more involved, due to the higher parton multiplicity and the fact that the mass of the bottom quark is treated exactly. 
Currently the inclusive cross section in the 4FS is only available through NLO~\cite{Dittmaier:2003ej,Dawson:2003kb,Wiesemann:2014ioa}.

It is known that perturbative results may differ substantially between both schemes, and various methods have been proposed to combine the 4FS and 5FS into a single prediction~\cite{Harlander:2011aa,Bonvini:2015pxa,Forte:2015hba,Bonvini:2016fgf,Forte:2016sja}.
A direct comparison of the NLO and NNLO results in the 4FS and 5FS respectively is, however, not straightforward, because they correspond to different orders in the perturbative expansion in the strong coupling constant (see Tab.~\ref{fig:diagrams}). 
A consistent comparison with the 4FS at NLO requires the knowledge of the inclusive cross section in the 5FS at next-to-next-to-next-to-leading order (N$^3$LO) in the strong coupling. 
In this paper we present for the first time the complete result for the inclusive cross section for Higgs production in bottom-quark fusion at N$^3$LO in the 5FS and investigate its phenomenological implications.

\section{The N$^3$LO cross section in the 5FS}
The inclusive hadronic cross section for the production of a Higgs boson can be written as
\beq
\label{eq:fac}
\sigma \!=\! \sum_{i,j}\!\int_0^1\!\! dx_1\,dx_2\,f_i(x_1,\mu_f)f_j(x_2,\mu_f)\hat{\sigma}_{ij}(z,\mu_r,\mu_f)\,,
\eeq
where the sum runs over all parton flavours, $f_i$ are parton densities and $\hat{\sigma}_{ij}$ are partonic cross sections. 
The partonic cross sections depend on the ratio $z=m_H^2/s$, where $\sqrt{s}$ is the partonic centre-of-mass energy, related to the hadronic centre-of-mass energy $\sqrt{S}$ by $s=x_1x_2S$ through the two Bjorken momentum fractions $x_{1,2}$. $\mu_r$ and $\mu_f$ denote the renormalisation and factorisation scales respectively. 
We work in the 5FS with five massless quark flavours. 
We assume that the Higgs boson has a non-zero Yukawa coupling $y_b$ to the bottom quark, and we neglect couplings of the Higgs boson to all other quark flavours. 
This implies that we focus on terms in the cross section proportional to $y_b^2$. 
The partonic cross sections $\hat{\sigma}_{ij}$ are expanded through N$^3$LO in the strong coupling $\alpha_s$. The complete set of initial state configurations that contribute to the cross section through N$^3$LO are shown in Tab.~\ref{fig:diagrams}.
The NLO and NNLO corrections to the cross section have been computed in Refs.~\cite{Dicus:1998hs,Balazs:1998sb,Harlander:2003ai}. 
In the remainder of this paper we present for the first time N$^3$LO corrections.

In order to compute the partonic cross sections at N$^3$LO, we follow the same steps that have been employed in the computation of the N$^3$LO corrections to Higgs production in gluon fusion in Refs.~\cite{Anastasiou:2015ema,Anastasiou:2016cez,Mistlberger:2018etf}. 
We have generated all relevant Feynman diagrams with QGraf~\cite{qgraf}. 
Individual Feynamn diagrams are sorted into scalar integral topologies, which are then reduced to a set of master integrals via integration-by-parts identities~\cite{Chetyrkin:1981qh,Tkachov:1981wb} using an in-house code. 
Finally, the master integrals are computed analytically using the differential equations method~\cite{Kotikov:1990kg,Kotikov:1991hm,Kotikov:1991pm,Henn:2013pwa,Gehrmann:1999as}.
All the relevant master integrals are known analytically as a function of $z$ and have been evaluated in the context of the N$^3$LO corrections to the gluon-fusion cross section. 
In particular at N$^3$LO three-loop corrections to the Born process contribute, which have been computed for the first time in Ref.~\cite{Gehrmann:2014vha} using the master integrals computed in Refs.~\cite{Gehrmann:2006wg,Heinrich:2007at,Heinrich:2009be,Lee:2010cga,Baikov:2009bg,Gehrmann:2010ue,Gehrmann:2010tu} and the results of Ref.~\cite{Gehrmann:2014vha} agree with our computation. 
In addition, the N$^3$LO cross section receives contributions from partonic subprocesses involving fewer loops but additional real emissions in the final state. 
Single-real emission contributions from two-loop and squared one-loop diagrams have been considered in Ref.~\cite{Anastasiou:2013mca,Kilgore:2013gba,Duhr:2013msa,Li:2013lsa,Dulat:2014mda,Ahmed:2014pka}.
The master integrals for double-real virtual and triple-real contributions have been computed in Refs.~\cite{Anastasiou:2014vaa,Li:2014bfa,Li:2014afw,Anastasiou:2015yha,Anastasiou:2013srw,Anastasiou:2015ema} as an expansion around the production threshold  of the Higgs boson and exactly as a function of $z$ in Ref.~\cite{Mistlberger:2018etf}. 
Here we work exclusively with the master integrals of Ref.~\cite{Mistlberger:2018etf}.

Contributions from different initial states and/or parton multiplicities are individually ultraviolet (UV) and infrared (IR) divergent. 
We regulate the divergences by working in dimensional regularisation in $D=4-2\eps$ dimensions. 
UV divergences can be cancelled by replacing both the bare strong and Yukawa couplings by their renormalised values in the $\overline{\textrm{MS}}$-scheme. 
The UV-counterterm for the strong coupling constant has been determined through five loops in Refs.~\cite{Tarasov:1980au,Larin:1993tp,vanRitbergen:1997va,Baikov:2016tgj,Herzog:2017ohr}. 
The renormalisation constant for the Yukawa coupling is identical to the quark mass renormalisation constant of QCD in the $\overline{\textrm{MS}}$-scheme~\cite{Harlander:2003ai,vanRitbergen:1997va,Chetyrkin:1997dh,Czakon:2004bu,Baikov:2014qja}. 
IR divergences are absorbed into the definition of the PDFs using mass factorisation at N$^3$LO~\cite{Buehler:2013fha,Hoschele:2012xc,Hoeschele:2013gga}.
The mass factorisation involves convoluting lower-order partonic cross sections with the three-loop splitting functions of Refs.~\cite{Moch:2004pa,Vogt:2004mw,Ablinger:2017tan}. 
We have computed all the convolutions analytically in $z$ space using the {\sc PolyLogTools} package~\cite{Duhr:2019tlz}. 
We observe that all divergences cancel after UV renormalisation and mass factorisation. 
We emphasise that this is not only a strong cross check of our result, but, together with the results of Ref.~\cite{Anastasiou:2015ema} for gluon-initiated processes, this is the first time that the complete set of three-loop splitting functions of Refs.~\cite{Moch:2004pa,Vogt:2004mw} has been confirmed by an independent analytic computation. 
Moreover, this is the first time that the universality of QCD factorisation has been confirmed for hadron collisions for all partonic initial states. 
     \begin{figure}[!t]
\begin{center}
 \begin{subfigure}[b]{0.5\textwidth}
  \begin{flushright}
 \includegraphics[width=0.9587\textwidth]{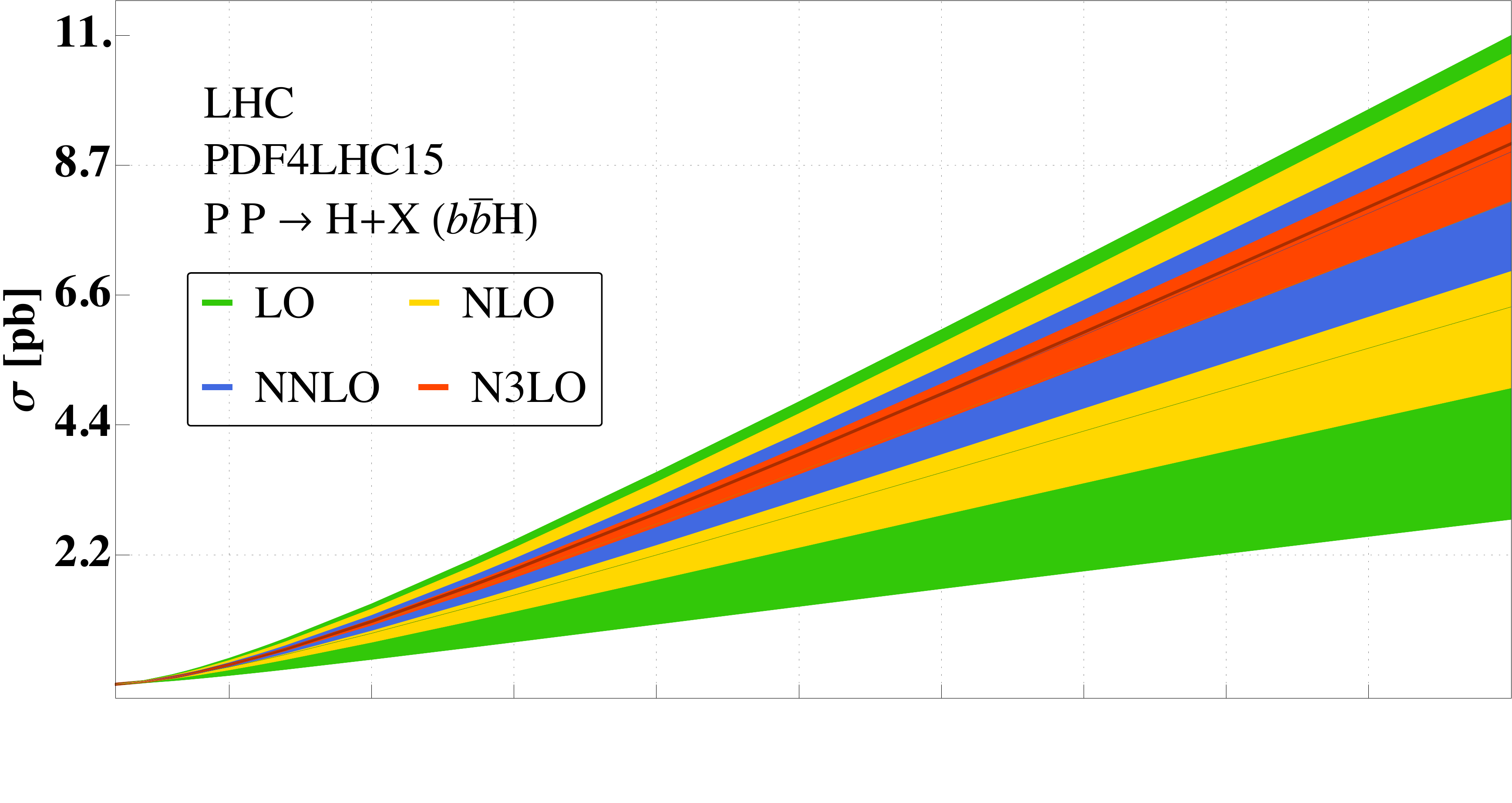}
\,\,
 \end{flushright}
\end{subfigure}
 \begin{subfigure}[b]{0.5\textwidth}
  \begin{flushright}
  \vspace{-0.5cm}
  \includegraphics[width=\textwidth]{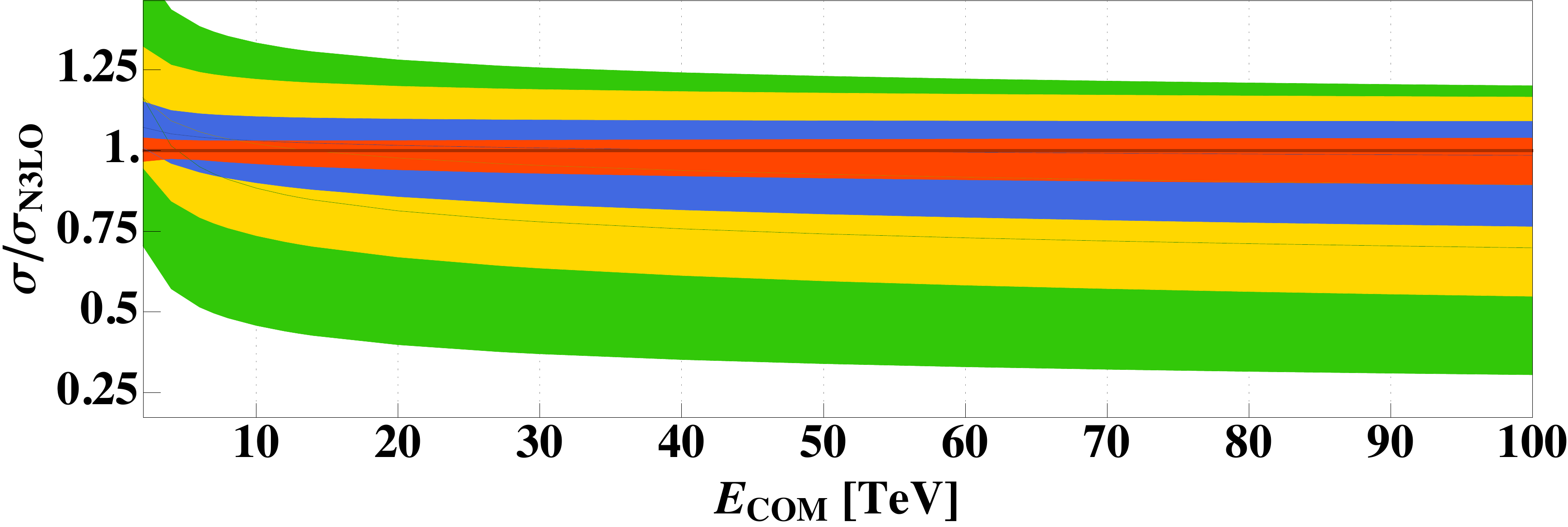}
 \end{flushright}
\end{subfigure}
  \caption{\label{fig:energy_variation}Variation of the hadronic cross section
      with the hadronic centre-of-mass energy. The upper figure shows nominal
      values, in the lower figure all predictions are normalised to the central value of the N$^3$LO prediction. 
      LO, NLO, NNLO and N$^3$LO corrections are shown in green, yellow, blue and red respectively. 
       The bands correspond to scale variation uncertainties as described in the text.}
  \end{center}
  \end{figure}
  
The analytic cancellation of all ultraviolet and infrared singularities provides a strong check of our results. 
In addition, we have reproduced the soft-virtual N$^3$LO cross section of Ref.~\cite{Ahmed:2014cha} and the physical kernel constraints of Ref.~\cite{Moch:2009hr,Soar:2009yh,deFlorian:2014vta} for the next-to-soft term of the bottom-quark-initiated cross section. 
We have also checked that all logarithmic terms in the renormalisation and factorisation scales produced from the cancellation of the UV and IR poles satisfy the DGLAP evolution equation. 
Finally, we have also recomputed the NLO and NNLO cross sections, and we have checked that through NNLO our results are in perfect agreement with the literature results implemented in the code {\sc Sushi}~\cite{Harlander:2012pb}.
Analytic results for the partonic coefficient functions will be presented in ref.~\cite{follow_up_paper}.

  \section{bottom-quark fusion at N$^3$LO in QCD}

  In this section we present our phenomenological results for inclusive cross section for bottom-quark fusion at N$^3$LO in QCD. 
  We assume a Higgs mass of $m_H=125.09\,$GeV. 
  The strong coupling is $\alpha_s(m_Z^2) = 0.118$ and is evolved to the renormalisation scale $\mu_r$ using the four-loop QCD beta function in the $\overline{\textrm{MS}}$-scheme assuming five massless quark flavours. 
  The Yukawa coupling between the Higgs boson and the bottom quark is proportional to the bottom-quark mass in the $\overline{\textrm{MS}}$-scheme, and we evolve it from $m_b(m_b)=4.18\,$GeV~\cite{Tanabashi:2018oca} to the same renormalisation scale $\mu_r$ using four-loop running~\cite{Baikov:2014qja}.

  Fig.~\ref{fig:energy_variation} shows the inclusive cross section at a proton-proton collider as a function of the hadronic centre-of-mass energy.
  The predictions are obtained by convoluting the partonic cross sections with
  the {\sc PDF4LHC15} NNLO PDFs in the
  5FS~\cite{Butterworth:2015oua}\footnote{
  It was pointed out in Ref.~\cite{Bonvini:2015pxa} that multiple different values for the bottom quark mass were used in the construction of the PDF4LHC15 sets and 
  an alternative PDF was derived. A PDF set where bottom mass effects are consistently included into the {\tt pdf4lhc\_nnlo\_mc} set is avilable from Ref.~\cite{Bonvinipage} (see also Ref.~\cite{deFlorian:2016spz}).
We find that using the PDF set of Ref.~\cite{Bonvinipage} introduces a $\mathcal{O}(1\%)$ shift of the central value of our cross section. 
Since the modification using the alternative PDF set is small we choose to use the official PDF4LHC15 sets of Ref.~\cite{Butterworth:2015oua} in our predictions for generality. 
For further discussion of bottom quark mass effects we refer to Ref.~\cite{follow_up_paper}. }
   as in eq.~\eqref{eq:fac}.
  The central value corresponds to the commonly used choice of the renormalisation and factorisation scales $(\mu_r,\mu_f)=(m_H,m_H/4)$ following for example refs.~\cite{Harlander:2003ai,Maltoni:2003pn}.
  The band is obtained by varying $\mu_r$ and  $\mu_f$ independently within the intervals $\mu_r\in[m_h ,2 m_h]$ and $\mu_f\in[m_h/8,m_h/2]$ with the restriction that $1/2\leq 4 \mu_f/\mu_r \leq 2$. 
  We observe that cross section predictions based on successive perturbative orders are contained within the bands of the lower order predictions over a wide range of hadronic centre of mass energies.
  The dependence on the renormalisation and factorisation scales of the hadronic cross section is reduced as the perturbative order is increased. 
  We therefore believe that the residual scale dependence provides a reliable  estimate of the missing higher orders beyond N$^3$LO.
Let us comment on the unconventionally small choice of the factorisation scale, $\mu_f=m_H/4$.
At NLO it was observed~\cite{Boos:2003yi,Plehn:2002vy,Rainwater:2002hm,Maltoni:2003pn}
that the $t$-channel singularity in the gluon-initiated process $gb\to bH$ leads to a collinear logarithm of the form $\log(4\mu_f/m_H)$ in the inclusive cross section. Based on this observation,
it was argued that is natural to vary to factorisation scale at NLO in an
interval around the central scale
$\mu_f=m_H/4$~\cite{Boos:2003yi,Plehn:2002vy,Rainwater:2002hm,Maltoni:2003pn}. This choice is corroborated by the fact that
it reduces the size of the NLO corrections. An improved convergence of the
perturbative expansion for $\mu_f=m_H/4$ was later also observed at
NNLO~\cite{Harlander:2003ai,Harlander:2012pb,deFlorian:2016spz}.
We see from our calculation that this trend continues at N$^3$LO. We observe that for higher values of $\mu_f$, the convergence behaviour of the cross section with the perturbative order deteriorates. 
In particular, choosing $\mu_f = \mu_r = m_H$ reduces the overlap of the scale
variation bands between NNLO and N$^3$LO.
We therefore conclude that $\mu_f=m_H/4$ remains a good choice for the
factorisation scale even at N$^3$LO.

Tab.~\ref{tab:xsecs} presents results for the inclusive cross section for a proton-collider for various hadronic centre-of-mass energies, and we show the main uncertainties affecting the cross section.
To obtain central values and PDF uncertainties we use the Monte-Carlo replica method following the PDF4LHC recommendation~\cite{Butterworth:2015oua}.
All other uncertainties are computed with a fixed PDF set, namely the zeroth member of the \texttt{PDF4LHC15\_nnlo\_mc} set.
The QCD scale uncertainty is estimated by varying $\mu_r$ and $\mu_f$ as described in the previous paragraph.
We also include an uncertainty reflecting the fact that currently there are no
N$^3$LO PDF sets available. 
The estimate of this uncertainty was obtained following the recipe introduced in Ref.~\cite{Anastasiou:2016cez}. 
The bottom quark mass is affected by an uncertainty of ${}^{+0.04}_{-0.03}$ GeV~\cite{Tanabashi:2018oca} and we display the resulting uncertainty of our cross section.
A more detailed discussion of the uncertainties, including a more comprehensive study of the impact of different PDF sets, will be presented in Ref.~\cite{follow_up_paper}.

  \begin{table}[!h]
\begin{center}
\begin{tabular}{r | c c c c c}
\hline\hline
S [TeV] & $\sigma_{bbH}$ [pb] & scale & PDF+$\alpha_s$  & $m_b$ & N$^3$LO PDFs \\
 \hline
 $ 7$ & $0.174 $ & ${}^{+3.0\%}_{-3.3 \%} $ & $ \pm 9.2 \% $ & $ {}^{+2.3 \%}_{-1.7\%} $ & $ \pm 3.9\% $ \\ \hline  
 $ 8$ & $0.226 $ & ${}^{+3.0\%}_{-3.6 \%} $ & $ \pm 9.2 \% $ & $ {}^{+2.3 \%}_{-1.7\%} $ & $ \pm 3.5\% $ \\ \hline  
 $ 13$ & $0.542 $ & ${}^{+3.0\%}_{-4.8 \%} $ & $ \pm 8.5 \% $ & $ {}^{+2.3 \%}_{-1.7\%} $ & $ \pm 2.5\% $ \\ \hline  
 $ 14$ & $0.614 $ & ${}^{+3.0\%}_{-5.0 \%} $ & $ \pm 8.5 \% $ & $ {}^{+2.3 \%}_{-1.7\%} $ & $ \pm 2.3\% $ \\ \hline  
 $ 27$ & $1.69 $ & ${}^{+3.2\%}_{-6.8 \%} $ & $ \pm 7.7 \% $ & $ {}^{+2.3 \%}_{-1.7\%} $ & $ \pm 1.2\% $ \\ \hline  
 $ 100$ & $9.20 $ & ${}^{+3.8\%}_{-11. \%} $ & $ \pm 6.8 \% $ & $ {}^{+2.3 \%}_{-1.7\%} $ & $ \pm 0.76\% $ \\ \hline  
\hline
\end{tabular}
\caption{\label{tab:xsecs}\centerlast The hadronic bottom-quark-fusion Higgs boson production cross section at various centre of mass energies. For a description of the uncertainties, see main text.}
\end{center}
\end{table}

 \section{Comparison to the Santander-matching}
 While the 5FS resums collinear logarithms to all orders, it neglects power-suppressed terms of the order $(m_b/Q)^2$, where $Q$ is a characteristic hard scale of the process. 
 For the inclusive cross section typically $Q\sim m_H$, so that $(m_b/Q)^2\sim 10^{-3}$, and the 5FS is expected to give reliable predictions. 
 Nevertheless, it has been observed that results in the 5FS at NNLO and in the 4FS at NLO may differ substantially. 
 However, this naive comparison is not satisfactory, because results in the 5FS at NNLO and in the 4FS at NLO correspond to different orders in perturbation theory. 
In particular, the 4FS at NLO includes terms proportional to $\alpha_s^3$ which are not captured by the 5FS at NNLO (see Tab.~\ref{fig:diagrams}).
For this reason various methods have been proposed to combine the 4FS and 5FS into a single prediction~\cite{Harlander:2011aa,Bonvini:2015pxa,Forte:2015hba,Bonvini:2016fgf,Forte:2016sja}. 
In this section we present for the first time a comparison of the two schemes that includes partonic cross sections consistently computed through $\alpha_s^3$.

The so-called Santander-matching (S-M) scheme~\cite{Harlander:2011aa} is widely used in the literature (see for example ref.~\cite{deFlorian:2016spz}) and the focus of this section.
It represents a pragmatic way of combining the 4FS and 5FS computations into a single prediction through a weighted average:
 \beq
 \sigma^{\textrm{S-M}}_{a,b} = \frac{\sigma^{4\textrm{FS,N$^a$LO}}+w\,\sigma^{5\textrm{FS,N$^b$LO}}}{1+w}\,,
 \eeq
 where the weighting factor is $w=\log\frac{m_H}{m_b}-2$. 
 
\begin{figure}[!t]
  \begin{center}
  \includegraphics[scale=0.32]{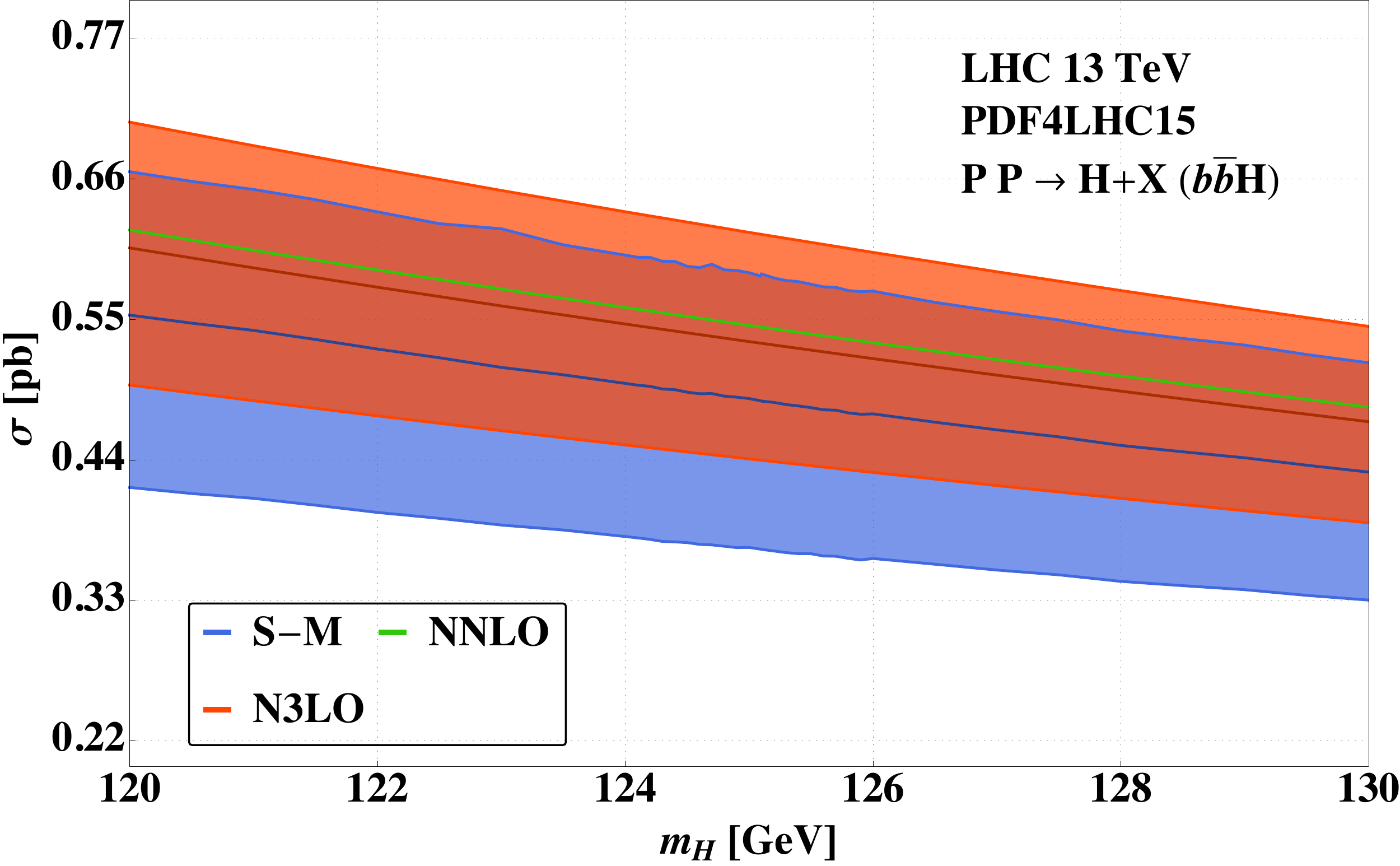}
  \caption{\label{fig:S-M}
  Comparison of $\sigma_{1,2}^{\textrm{S-M}}$ (blue) and the cross section computed in the 5FS at N$^3$LO (red) and NNLO (green).
  The bands represent a linear sum of PDF, scale-variation and bottom quark mass uncertainties. 
  For the 5 flavour scheme also an uncertainty for the miss-match of the PDF order and the order of the  partonic cross section is included.
  }
  \end{center}
  \end{figure}

 Fig.~\ref{fig:S-M} shows the inclusive cross section computed in the S-M scheme compared to the results in the 5FS at NNLO and N$^3$LO. 
 The values of the cross section in the S-M scheme for $(a,b)=(1,2)$ have been obtained from the reference values published in Ref.~\cite{deFlorian:2016spz} based on Refs.~\cite{Harlander:2012pb,Wiesemann:2014ioa,Bonvini:2016fgf}. 
 We see that the N$^3$LO corrections lower the value of the cross section in the 5FS, bringing it closer to the results in the S-M scheme. 
 Moreover, we see that our N$^3$LO results and the S-M results agree within scale and PDF uncertainties. Since the S-M scheme is a pragmatic, but very ad hoc, prescription, we do not see any compelling argument to work with this scheme, given that now the 4FS and 5FS are known to the same order in the strong coupling constant and that power-suppressed terms in the 4FS are expected to give small contributions to the inclusive cross section.

 \section{Conclusions}
 We have presented for the first time the complete computation of the inclusive cross section for the production of a Higgs boson through bottom-quark fusion at N$^3$LO. 
 We observe a substantial reduction of the residual scale uncertainty and a good
 convergence of the perturbative series, provided that the factorisation scale
 is set to a small value around $m_H/4$, in agreement with previous studies. We have also compared the value of the inclusive cross section at N$^3$LO in the 5FS to the S-M of Ref.~\cite{deFlorian:2016spz}. 
 We find that the two results agree within uncertainties. 
Since finite power-suppressed effects due to the bottom-mass are expected to give only small contributions to the cross section, we believe that our result in the 5FS provides the most reliable prediction of the perturbative QCD corrections to the inclusive cross section of Higgs production to date, including consistently all higher-order contributions through $\alpha_s^3$.

\begin{acknowledgments}
\emph{Acknowledgements:}  We are grateful to  Markus Ebert, Einan Gardi, Davide Napoletano,
Iain Stewart and Lorenzo Tancredi for discussion, and to Valentin Hirschi for
help with installing \verb+MadGraph5_aMC@NLO+.
Feynman diagrams have been drawn with the Feynman-Tikz package~\cite{Ellis:2016jkw}.
The research of CD is supported
by the ERC grant 637019 ``MathAm''. BM is supported by the Pappalardo
fellowship, and the research of FD is supported by the  U.S.  Department  of Energy (DOE) under contract DE-AC02-76SF00515.
\end{acknowledgments}

\bibliography{main}

\end{document}